# Resonant Asymmetric All-Dielectric Metasurface for Boosting Third-Harmonic Generation

Chaobiao Zhou,[†,§,⊥] Shiyu Li,[†] Cheng Gong,[§] Yi Wang,[*,†] Xiaojun Liu,[*,§] and Mingsheng Zhan[*,†,§]

[†]*Wuhan National Laboratory for Optoelectronics, Huazhong University of Science and Technology, Wuhan, Hubei, 430074, China*

[§]*State Key Laboratory of Magnetic Resonance and Atomic and Molecular Physics, Wuhan Institute of Physics and Mathematics, Chinese Academy of Sciences, Wuhan, Hubei, 430071, China*

[⊥]*College of Mechanical and Electronic Engineering, Guizhou Minzu University, Guiyang 550025, China*

**Abstract:** Resonant metasurfaces have received extensive attention due to their sharp spectral feature and extraordinary field enhancement. In this work, by breaking the in-plane symmetry of silicon nanopillars, we achieve a sharp Fano resonance. The far-field radiation and near-field distribution of metasurfaces are calculated and analyzed to further uncover the resonant performance of metasurfaces. Moreover, the theoretical derivation and simulation exhibit an inverse quadratic dependence of Q-factors on asymmetry parameters, revealing that the resonance is governed by the symmetry-protected bound states in the continuum. Finally we experimentally demonstrate the sharp resonance, and employ it to effciently boost the third-harmonic generation. This enhancement can be attributed to the strong optical intensity enhancement inside the metasurface.

Optical metasurfaces, the planar elements of three dimensional metamaterials, have become a rapidly growing research field in recent years owing to their unique properties in light manipulation.[1-3] Fano resonance (FR), originating from the interference between discrete and continuous states, can be achieved by appropriately choosing materials and metasurface structures.[4-6] The sharp spectral features and strong field enhancements of FR promise applications in light emission,[7] biosensing,[8] optical switching,[9] slow light,[10] nonlinear optics[11] and so on. As well known, the early FR metasurfaces, consisting of dielectric layer and metal film, exhibit particularly broad resonances in the optical range due to high inherent energy dissipation of metal.[12] In order to reduce dissipative losses and enhance electromagnetic field intensity in FR metasurface, a variety of all-dielectric nanostructures with low losses have been developed.[13-18] It should be noted that, for the dielectric nanostructures, the local field enhancement occurs inside the metafurface. This is different from their plasmonic counterparts, for which the electric field is confined in the surface regions. Recently, a number of all-dielectric metasurfaces with symmetry breaking have been designed to achieve strong interference between the "bright" and "dark" collective modes, resulting in the sharp FR with high quality factor (Q-factor).[19-23] Therein, the bright mode refers to the broadband continuum state and it is easily coupled with incident light, the dark mode corresponds to the discrete resonance state which exhibits a weak-coupling with incident light unless the symmetry of the structure is broken. The high Q-factor FR could provide a powerful approach to facilitate the light manipulation at the nanoscale, especially in the nonlinear regime.[24-25]

Third-harmonic generation (THG) is one of important optical nonlinear processes. Efficient THG is beneficial for a variety of applications including bioimaging techniques[26-27] and pulse characterzations.[28] Over the past years, many plasmonic nanostructures, such as nanoparticles,[29] bowties,[30] fishnets,[31] nanostripes[32] and dolmen-types,[33] have been explored to enhance the THG in the optical range. Furthermore, both theoretical and experimental investigations suggested that a few metal-dielectric hybrid nanostructures may be used to reduce THG loss.[34-36] However, the enhancement of THG based on metal nanostructures still suffers from an intrinsic limitation, which arises from the Ohmic losses in the optical range.[12] Recently, all-dielectric silicon nanostructures have been extensively considered to enhance THG, owing to their CMOS compatibility, strong optical nonlinearities and almost negligible optical losses in the near-infrared region.[37] Those dielectric nanostructures include silicon nanodisks,[38-41] photonic crystal waveguides,[28, 42] oligomers[43] and other nanostructures.[44-50] However, the aforementioned silicon nanostructures still display strong leaky characteristics of the optical modes, which limits the enhancement of THG. Thus, it is still in demand to seek higher Q-factor resonator which will further facilitate the manipulation of the nonlinear interaction of light.

In this work, we experimentally achieve a high Q-factor FR in a silicon metasurface consisting of asymmetric nanopillar arrays, owing to its low radiation loss, large fabrication tolerance and coherent interaction of the neighbouring asymmetric nanopillars. We also experimentally demonstrate the giant THG from the Fano resonant silicon metasurface and a maximal enhancement factor of about 600. In addition, the polarization property of THG is studied in detail.

**Resonant metasurface design, fabrication and discussion**

The resonant silicon metasurface consists of an array of nanopillars with a symmetry-breaking geometric structure, as schematically shown in Figure 1a. The sample is designed on a silicon-on-insulator (SOI) substrate, and fabricated through electron-beam lithography (EBL) and inductively coupled plasma (ICP) etching techniques. The scanning electron microscope (SEM) images of the fabricated sample are shown in Figure 1b.

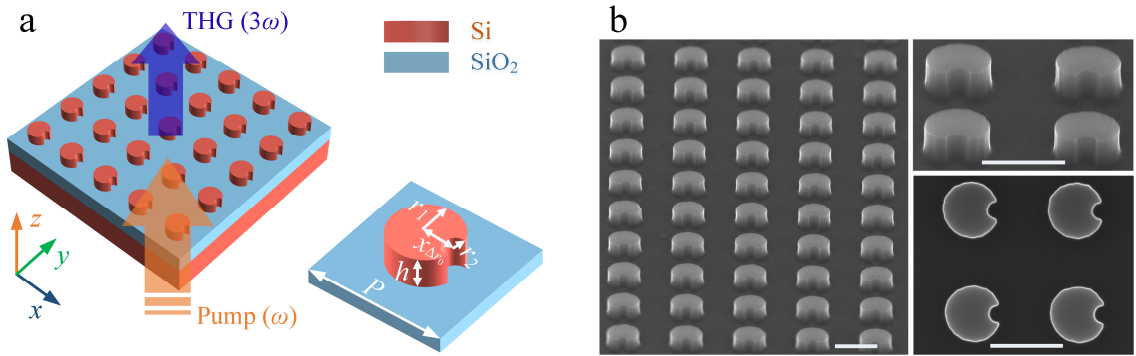

**Figure 1.** THG process in resonant silicon metasurface. a) Illustration of THG from metasurface with the magnetic dipole resonance. The pulse with fundamental frequency $\omega$ (energy $E = \hbar\omega$) is coupled to the backside of silicon metasurface, and the third-harmonic signal (frequency $\omega_{THG} = 3\omega$) is generated from metasurface along $z$-axis. In the inset, the geometrical parameters of the asymmetrical nanopillar are given: the unit cell period $P = 800$ nm, nanopillar height $h = 220$ nm and nanopillar radius $r_1 = 200$ nm, respectively. The air-hole radius $r_2 = 70$ nm, and the center distance between the nanopillar and the air-hole $x_{\Delta r_0} = 180$ nm. b) Scanning electron microscope images from oblique view (left), oblique and top views (right) after zooming in on four asymmetrical nanopillars of fabricated metasurface. The scale bars correspond to 500 nm.

Finite-difference-time-domain (FDTD) method is employed to calculate the optical properties of metasurface. As illustrated in Figure 2a, for the case of the symmetric metasurface with electric field parallel to $y$-axis ($E \parallel y$), the uniform transmission is observed in a wide wavelength range (black line). Interestingly, we can obtain the sharp FR profile in the asymmetric metasurface with $E \parallel y$ in Figure 2b. The resonance wavelength is 1341 nm. It is noted that the FR is sensitive to the polarization of the electromagnetic field. When the electric field of incident beam is parallel to the $x$-axis, the resonance disappears as illustrated in Figure 2a (red line). The results reveal that strong FR can be achieved by introducing the symmetry-breaking in all-dielectric nanopillars and pumping them with a normal-incident plane beam with the electric field oriented along an appropriate polarization angle. In the following, the physical mechanism of sharp FR is discussed. For nanopillars possessing high-symmetry geometry, the electric dipole mode ($P_y$) oscillated along *y*-direction is excited under a *y*-polarized electromagnetic wave, which is subject to both radiative and nonradiative decay processes.[19]

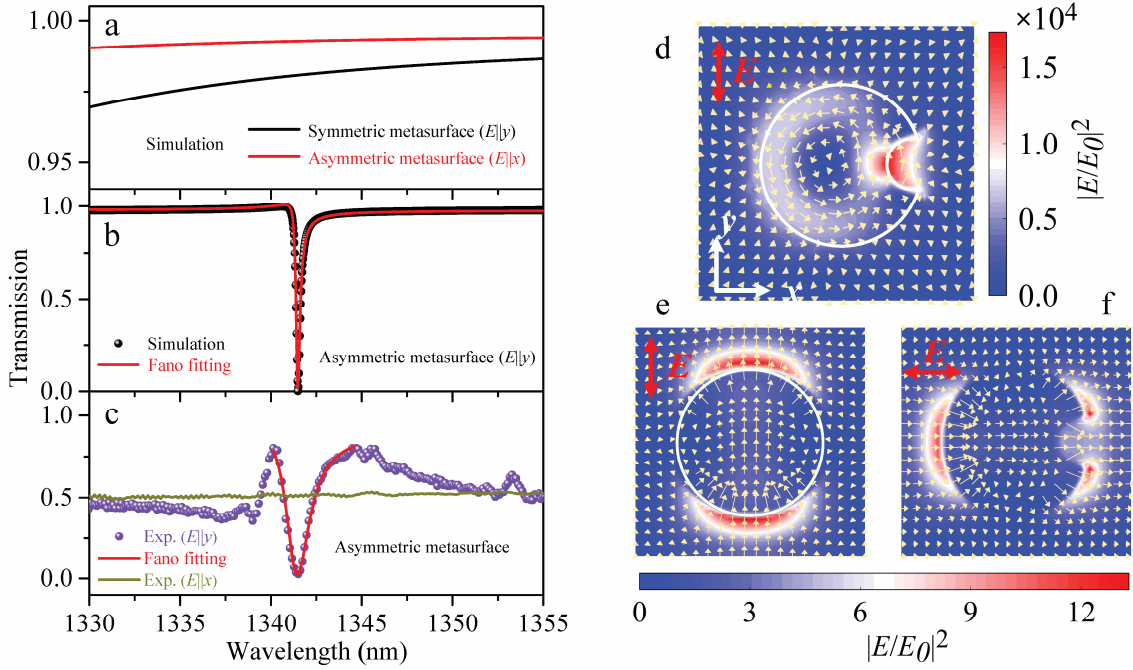

**Figure 2.** Theoretical and experimental transmission spectra, and simulated electric field distributions. a,b) The simulated transmission spectra are calculated by FDTD method, three cases are discussed including symmetric metasurface with incident electric field parallel to *y*-axis ( $E \parallel y$ ), asymmetry metasurface with incident electric field parallel to *y*-axis ( $E \parallel y$ ) and *x*-axis ( $E \parallel x$ ). The sharp FR at wavelength $\lambda_{Res}$= 1341 nm is excited when introducing the symmetry-breaking in nanopillar with $E \parallel y$ . c) Experimental transmission spectra of asymmetric metasurface for $E \parallel y$ and $E \parallel x$ . The simulated and experimental resonance peaks in (b), (c) are fitted by Fano formula (red line). d-f) The simulated electric field patterns in *x-y* plane at the center of nanopillar at wavelength 1341 nm. They correspond to the three cases in (a), (b), respectively. White solid line denotes the boundary of silicon region. The arrow represents the direction of circular displacement current, the color scale corresponds to the field intensity. Strong local field enhancement exists for the case of asymmetric metasurface with $E \parallel y$ due to the magnetic dipole oscillation in the nanopillar center.

This results in collective oscillations of the nanopillars forming the bright mode (black line in Figure 2a). After cutting a notch at the edge of the nanopillar, the symmetry-breaking in structure is introduced, the electric dipoles $P_y$ will exhibit slightly different dipole strengths in the notch part and opposite side. The asymmetrical two dipoles will generate a *z*-directed magnetic field near the center of the nanopillar which will excite the $M_z$ dipole. It serves as a dark mode because near-field coupling of nanopillars dramatically suppresses the radiative loss.

Thus, the symmetry breaking induces interference between the bright in-plane electric dipole mode $P_y$ and the dark magnetic dipole mode $M_z$, leading to the observed high Q-factor FR.[7, 19]

In our experiment, a sufficiently large metasurface (640 μm×640 μm) is fabricated in order to reduce strong scattering of light into free-space.[19, 51-52] The measured transmission spectra of the asymmetric metasurface for $E \parallel y$ and $E \parallel x$ are shown in Figure 2c. We fit the Fano lineshape by the following classical Fano formula:[53-55]

$$T(\omega) = T_0 + A_0 \frac{[q + 2(\omega - \omega_0)/\Gamma]^2}{1 + [2(\omega - \omega_0)/\Gamma]^2}, \quad (1)$$

where $\omega_0$ is the resonant frequency, $\Gamma$ is the resonance linewidth, and $T_0$ is the transmission offset, $A_0$ is the continuum-discrete coupling constant, $q$ is the Breit-Wigner-Fano parameter determining asymmetry of the resonance profile. The Q-factor is evaluated by $\frac{\omega_0}{\Gamma}$.

The fitted results are shown in Figure 2b,c (red lines). The experimental transmission spectrum is only fitted in the wavelength range of resonance peak, because it is difficult to obtain a standard Fano lineshape as simulation for a fabricated all-dielectric metasurface due to fabrication deviation.[20, 22, 51] Herein, The evaluated Q-factors are about 5300 and 1000 in simulation and experiment, respectively. The experimental value is lower than that in theory, which could be mainly attributed to incoherent scattering from the fabrication imperfections, such as the surface roughness and nonuniformities of the periodic nanopillars.[51, 56] Our design holds a relatively large fabrication tolerance, comparing with the composite structures shown in Refs. [22,23,51], this kind of metasurface whose unit cell has only single nanostructure, it greatly eases fabrication challenges since it does not require the production of deep subwavelength gaps or spaces between resonators.[19] Comparing to nanostructures with only one dielectric structure per unit cell, such as the notched cube shown in Refs. [7,19], our design holds fewer sharp corners, it is much more accessible in experiment fabrication, shuch as EBL and ICP procedures *etc*.

The simulated electric field distributions in *x-y* plane at the wavelength of 1341 nm are shown in Figure 2d-f, corresponding to the three cases in Figure 2a,b. As illustrated in Figure 2d, the maximum of local electric field enhancement is $|E/E_0|^2 = 1.9 \times 10^4$, where $E$ and $E_0$ are the amplitudes of local electric field and incident electric field, respectively. The polarization of the electric field is anti-parallel at opposite sides of the breaking nanopillar induced by the varied effective refractive index of silicon metasurface. This results in a distinct circular displacement current inside the nanopillar in the *x-y* plane, revealing that the energy within the metasurface is strongly trapped by the magnetic dipole moment oscillation along z axis in the center of each nanopillar (depicted in Figure S2). As for the symmmetric nanopillars with $E \parallel y$ or asymmetric nanopillar with $E \parallel x$, the magnetic dipole resonance vanishes, as shown in Figure 2e,f, resulting in a maximum value of $|E/E_0|^2$ of only about 13.

To further confirm the role of magnetic dipole contribution forming the sharp FR, the electromagnetic multipole expansion is performed. By adopting Cartesian coordinate, the multipole moments can be obtained based on current density $\vec{j}$ :[57]

$$\vec{P} = \frac{1}{i\omega} \int \vec{j} d^3 r, \tag{2}$$

$$\vec{M} = \frac{1}{2c} \int (\vec{r} \times \vec{j}) d^3 r, \tag{3}$$

$$\vec{T} = \frac{1}{10c} \int [(\vec{r} \cdot \vec{j})\vec{r} - 2r^2 \vec{j}] d^3 r, \tag{4}$$

$$Q^{(e)}_{\alpha,\beta} = \frac{1}{2i\omega} \int [r_\alpha j_\beta + r_\beta j_\alpha - \frac{2}{3}(\vec{r} \cdot \vec{j})\delta_{\alpha,\beta}] d^3 r, \tag{5}$$

$$Q^{(m)}_{\alpha,\beta} = \frac{1}{3c} \int [(\vec{r} \times \vec{j})_\alpha r_\beta + ((\vec{r} \times \vec{j})_\beta r_\alpha)] d^3 r, \tag{6}$$

where $c$ is the speed of light, $\omega$ is the frequency of light and $\alpha, \beta, \gamma = x, y, z$. The $\vec{P}, \vec{M}, \vec{T}$, $Q^{(e)}, Q^{(m)}$ are electric dipole moment, magnetic dipole moment, toroidal dipole moment, electric quadrupole moment and magnetic quadrupole moment, respectively. The total scattered power of the multipole moments is calculated from the following formula:[57]

$$I = \frac{2\omega^4}{3c^3}|\vec{P}|^2 + \frac{2\omega^4}{3c^3}|\vec{M}|^2 + \frac{4\omega^5}{3c^4}(\vec{P} \cdot \vec{T}) + \frac{2\omega^6}{3c^5}|\vec{T}|^2 + \\ \frac{\omega^6}{5c^5}\sum |Q^{(e)}_{\alpha,\beta}|^2 + \frac{\omega^6}{40c^5}\sum |Q^{(m)}_{\alpha,\beta}|^2 + O\left(\frac{1}{c^5}\right). \tag{7}$$

The calculated results are shown in Figure 3, revealing the electric dipole (P), toroidal dipole (T), electric quadrupole ($Q^{(e)}$) and magnetic quadrupole ($Q^{(m)}$) only make a small contribution for the resonance at wavelength $\lambda = 1341$ nm. The main contribution is dominated by the magnetic dipole (M). The electromagnetic multipole expansion offers a theoretical proof for the simulation results of Figure 2d and Figure S2.

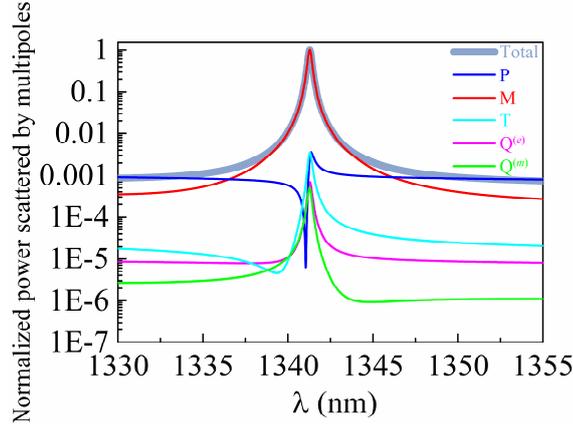

**Figure 3.** The normalized total scattering power (Total) and the contributions of electric dipole (*P*), magnetic dipole (*M*), toroidal dipole (*T*), electric quadrupole ($Q^{(e)}$) and magnetic quadrupole ($Q^{(m)}$). The log scale in the *y*-axis is chosen so as to reveal more clearly the contribution of every electromagnetic dipole.

Next, we study the dependence of the Q-factor on the asymmetry parameter α, which is defined as $\alpha = \frac{\Delta S}{S}$, as shown in the inset of Figure 4, $\Delta S$ is the cut-out area of the metasurface, and *S* is the area of symmetry nanopillar. In Fig.4, it is clear that the calculated results from FDTD satisfy the following relationship:

$$Q \propto \alpha^{-2}, \qquad (8)$$

which is consistent with the theory $Q = \frac{2A}{k_0}|p_0|^{-2}\alpha^{-2}$ (see part 1 of the Supporting Information), the asymmetric metasurfaces with high Q-factor resonances are verified to be governed by bound states in the continuum (BIC).[50, 58-59] Theoretically, ultra-high Q-factor can be achieved by decreasing the asymmetry parameter. However, it is difficult to achieve the ultra-high experiment Q-factor due to exacting fabrication techniques.[19, 22, 60]

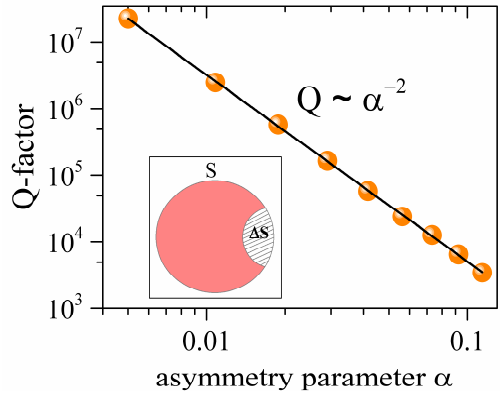

**Figure 4.** The Q-factor as a function of the asymmetric parameter α (log-log scale). The orange balls represent the simulated results, and the black line exhibits the inverse quadratic relationship between the Q-factor and α. The insert shows the specific shape of the isolated asymmetric nanopillar.

## Characterization of THG in resonant silicon metasurface

It is known that the conversion efficiency of THG correlates with the nonlinear polarization $P^{(3)}$, which strongly depends on the intensity of local electric field $E$ and the third-order nonlinear susceptibility of the material $\chi^{(3)}$. Because the unit cell of metasurface is smaller than the wavelength of incident light, strict phase matching is not taken into account.[32, 44, 61] Thus, the $P^{(3)}$ can be given as:

$$P^{(3)} \propto \int_V \chi^{(3)}(r) E^3(r,\omega) \mathrm{d}V , \qquad (8)$$

where $E(r,\omega)$ is the local electric field, $V$ is the volume of a unit cell. Our Fano resonant silicon metasurface possesses a high Q-factor and exhibits a strong local field enhancement as verified above. Furthermore, the silicon holds large third-order nonlinear susceptibility in the near-infrared region ($\chi^{(3)}_{Si} \approx 2.79 \times 10^{-18} m^2 \cdot V^{-2}$).[44] In the following, we experimentally demonstrate the strong THG from the resonant metasurface, which is performed by a home-built optics system as shown in Figure 5a (details seen Method). The measured THG spectra are shown in Figure 5b. In contrast with the unpatterned silicon film of the same thickness, the significant enhanced THG from metasurface is clearly observed with a maximum enhancement factor up to about 600 at upconverted wavelength $\lambda_{THG} = 447$ nm (enhancement factor is defined as the THG intensity ratio of metasurface and bulk film[38, 46]). This result can be attributed to the strong local electric field and relatively large mode volume inside the metasurface at the resonance wavelength $\lambda_{Res} = 1341$ nm.

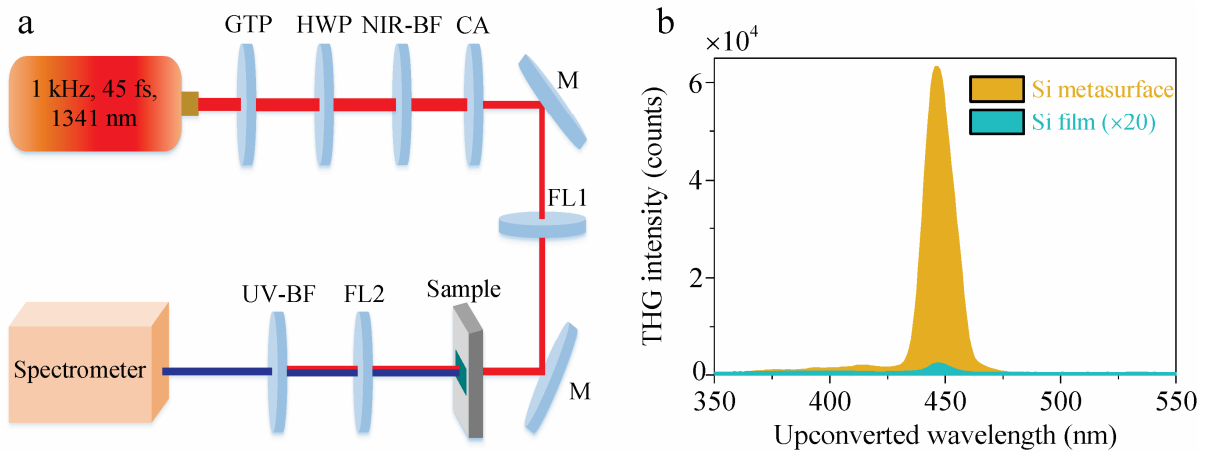

**Figure 5.** THG measured from a silicon metasurface. a) Schematic diagram of experimental setup for THG measurement. The optical components are listed below: Glan-Taylor polarizer (GTP), Broadband half-wave plate (HWP), NIR Bandpass Filter (NIR-BF, 700 -1650 nm), circular aperture (CA), mirror (M), self-focusing lens (FL1, FL2), UV/Visible Bandpass Filter (UV-BF, 340 - 694.3 nm) and optical spectrometer (Ocean Optics Inc., QE65PRO, 347 nm-1100 nm). b) The upconversion THG spectra measured for the unpatterned silicon film (×20

times) and silicon metasurface. The pump power of the infrared femtosecond laser pulses polarized along *y*-axis is 15 μW at central wavelength 1341 nm.

Note that, for those nanophotonics devices whose resonances hold strong polarization dependence, the harmonic beam intensity can be controlled by altering the polarization of the incident light.[62-63] In Figure 6, we give the polarization dependence of THG from the unpatterned silicon film and metasurface. For the unpatterned silicon film, it is found that THG intensity doesn't show obvious variations with different laser polarizations, as illustrated in Figure 6a. The reasons could be ascribed to two aspects: On the one hand, the silicon material with (100) crystal direction is used in our experiment, and it is anisotropic.[40] While the polarization feature of THG intensity from unpatterned silicon film is difficult to detect, because of the unstable femtosecond laser power and relatively weak THG signal from the silicon film. On the other hand, the silicon film doesn't exhibit a polarization-dependent resonance effect and, as a consequence, it just produces an unpolarized THG. In contrast, for metasurface, the THG intensity is very sensitive to the incident polarization, as shown in Figure 6b, where 0° or 180° corresponds to $E \parallel y$, and 90° or 270° refer to $E \parallel x$. It can be ascribed to the strong polarization dependence of the sharp FR, which is excited only when the electric field of the incident light is parallel to *y*-axis (as shown in Figure 2). It is well known that THG intensity is proportional to FR intensity. Consequently, THG intensity also displays a distinct polarization dependence. This effect could be used to achieve polarization manipulation of third-harmonic beam. For example, THG with circular polarization can be obtained by employing a circularly polarized light to pump two orthogonal arrays of metasurfaces.[62, 64]

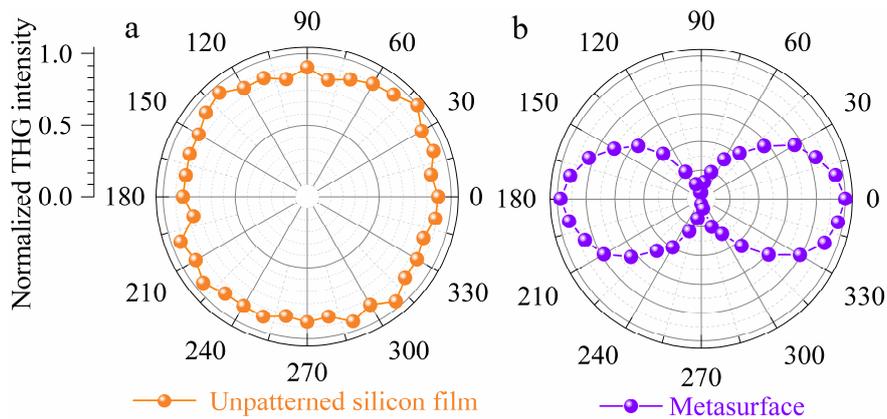

**Figure 6.** Polar plots of THG intensity as a function of the excitation laser polarization, THG for (a) the unpatterned silicon film, and (b) the asymmetrical silicon metasurface, respectively.

**Conclusions**

In summary, by introducing the symmetry-breaking to the highly symmetric silicon nanopillars, we have experimentally achieved a high Q-factor FR in the all-dielectric metasurface. This is contributed to the low radiation loss, large fabrication tolerance and effective coherent interaction of the metasurface. We have also demonstrated the THG from such a metasurface pumped by femtosecond laser pulses, and an enhancement factor of ~600 is obtained. This strong enhancement can be ascribed to the enhanced local field and relatively large mode volume inside the metasurface. Our results offer a route to achieve efficient and compact nonlinear light sources.

**Methods:**

*Electromagnetic Simulations:* The FDTD method is carried out to calculate the transmission spectra and the corresponding near-field distributions. In the simulation setup, the periodical boundary conditions are set in the *x* and *y* directions and perfectly matched layers are set in the *z* direction. The dielectric constants of Si and $SiO_2$ are directly adotpted from Palik.[65]

*Sample Fabrication:* The metasurface is fabricated on a SOI substrate with 220 nm top Si layer, 2 μm buried oxide layer and about 700 μm Si substrate. The fabrication processes include EBL and ICP etching techniques. Firstly, a clean SOI flake is spin-coated with 340 nm ZEP-520A as a resist. Then the nanopillar array patterns are defined into the resist by EBL (Vistec EBPG 5000 Plus), the beam current is about 0.5 nA, the exposure dose is 210 μC/cm$^2$ and the beam step size is 4 nm. After being carefully developed and fixed, the patterns are transferred to the top Si by ICP-RIE (Oxford Plasmalab System 100 ICP180) etching technique using a *$SF_6$/$C_4F_8$* mixture. The remaining resist is ultrasonically removed by dipping into the N-methyl-2-pyrrolidone (NMP) liquor.

*Transmission spectroscopy measurements:* The transmission spectra of the fabricated metasurface are measured by a home-built setup, where the input light from a picosecond pulsing laser (YSL Photonics SC-5-FC, 480-2200 nm) is incident onto the metasurface with a fiber collimator. The size of the fabricated metasurface is large enough to ensure that the whole light beam is passing through the patterned area. The light transmitted through is then collected on the backside of the sample using another fiber collimator and directed to a spectrometer (YOKOGAWA AQ-6370).

*THG measurements:* THG is measured by a home-built optics system. The pump light pulses centered at $\lambda_{pump}$= 1341 nm are generated via an optical parametric amplifier (OPA) system pumped by a Ti: Sapphire femtosecond laser (Coherent, Legend Elite HE + USP-1K-III, the pulse duration is 45 fs and repetition rate is 1 kHz, the pump wavelength can be tunable from 0.2 μm to 2.6 μm). The laser power is regulated by a Glan-Taylor Polarizer (GTP). In order to change the polarization angle of electric field of the pump light, a half-wave plate is placed behind the GTP. The visible light accompanied with the pump beam can be filtered by an NIR Bandpass Filter. A circular aperture is adopted to shrink the size of the pump light, and then a focus lens is used to gather the laser power. The pump beam is steered to the backside of the silicon metasurface to avoid the strong absorption of THG in bulk silicon substrate.[54,56][43, 45, 66] Another focus lens is placed after the sample to collect THG signal. Then a UV/Visible Bandpass Filter is utilized to eliminate the near-infrared light from the pump beam. Finally, the spectrometer is employed to record the THG signal.

**Supporting Information**

The Supporting Information is available free of charge on the ACS Publications website at DOI: XXXX

Derivation of the dependence of the Q-factor on the asymmetric parameter α. Magnetic field distribution in *x-z* plane. Power dependence of THG in metasurface. Diffraction of THG in the resonant metasurface.

**Corresponding Author**
*E-mail: ywangwnlo@mail.hust.edu.cn.
*E-mail: xjliu@wipm.ac.cn.
*E-mail: mszhan@wipm.ac.cn.

**Notes**
The authors declare no competing financial interest.

**Acknowledgment**
The authors thank all engineers in the Center of Micro Fabrication and Characterization (CMCF) of Wuhan National Laboratory for Optoelectronics (WNLO) for the support in fabrication, and would also like to thank Dr. Shuyuan Xiao (Nanchang University) for fruitful discussions.This work was supported by the National Natural Science Foundation of China (Grant No. 61775075).

**References**

(1) Kildishev, A. V.; Boltasseva, A.; Shalaev, V. M. Planar photonics with metasurfaces. *Science* **2013**, *339* (6125), 1232009.

(2) Hsiao, H.-H.; Chu, C. H.; Tsai, D. P. Fundamentals and applications of metasurfaces. *Small Methods* **2017**, *1* (4), 1600064.

(3) Boardman, A. D.; Grimalsky, V. V.; Kivshar, Y. S.; Koshevaya, S. V.; Lapine, M.; Litchinitser, N. M.; Malnev, V. N.; Noginov, M.; Rapoport, Y. G.; Shalaev, V. M. Active and tunable metamaterials. *Laser & Photonics Reviews* **2011**, *5* (2), 287-307.

(4) Fano, U. Effects of configuration interaction on intensities and phase shifts. *Phy. Rev.* **1961**, *124* (6), 1866-1878.

(5) Miroshnichenko, A. E.; Flach, S.; Kivshar, Y. S. Fano resonances in nanoscale structures. *Rev. Mod. Phys.* **2010**, *82* (3), 2257-2298.

(6) Limonov, M. F.; Rybin, M. V.; Poddubny, A. N.; Kivshar, Y. S. Fano resonances in photonics. *Nat. Photonics* **2017**, *11* (9), 543-554.

(7) Liu, S.; Vaskin, A.; Addamane, S.; Leung, B.; Tsai, M. C.; Yang, Y.; Vabishchevich, P. P.; Keeler, G. A.; Wang, G.; He, X.; Kim, Y.; Hartmann, N. F.; Htoon, H.; Doorn, S. K.; Zilk, M.; Pertsch, T.; Balakrishnan, G.; Sinclair, M. B.; Staude, I.; Brener, I. Light-emitting metasurfaces: Simultaneous control of spontaneous emission and far-field radiation. *Nano Lett.* **2018**, *18* (11), 6906-6914.

(8) Wu, C.; Khanikaev, A. B.; Adato, R.; Arju, N.; Yanik, A. A.; Altug, H.; Shvets, G. Fano-resonant asymmetric metamaterials for ultrasensitive spectroscopy and identification of molecular monolayers. *Nat. Mater.* **2011**, *11* (1), 69-75.

(9) Manjappa, M.; Srivastava, Y. K.; Solanki, A.; Kumar, A.; Sum, T. C.; Singh, R. Hybrid lead halide perovskites for ultrasensitive photoactive switching in terahertz metamaterial devices. *Adv. Mater.* **2017**, *29* (32).

(10) Wu, C.; Khanikaev, A. B.; Shvets, G. Broadband slow light metamaterial based on a double-continuum fano resonance. *Phys. Rev. Lett.* **2011**, *106* (10), 107403.

(11) Liu, S. D.; Leong, E. S. P.; Li, G. C.; Hou, Y. D.; Deng, J.; Teng, J. H.; Ong, H. C.; Lei, D. Y. Polarization-independent multiple fano resonances in plasmonic nonamers for multimode-matching enhanced multiband second-harmonic generation. *ACS Nano* **2016**, *10* (1), 1442-1453.

(12) Khurgin, J. B. How to deal with the loss in plasmonics and metamaterials. *Nat. Nanotechnol* **2015**, *10* (1), 2-6.

(13) Miroshnichenko, A. E.; Kivshar, Y. S. Fano resonances in all-dielectric oligomers. *Nano Lett.* **2012**, *12* (12), 6459-63.

(14) Huang, L.; Yu, Y.; Cao, L. General modal properties of optical resonances in subwavelength nonspherical dielectric structures. *Nano Lett.* **2013**, *13* (8), 3559-65.

(15) Kuznetsov, A. I.; Miroshnichenko, A. E.; Brongersma, M. L.; Kivshar, Y. S.; Luk'yanchuk, B. Optically resonant dielectric nanostructures. *Science* **2016**, *354* (6314), aag2472.

(16) Tuz, V. R.; Khardikov, V. V.; Kivshar, Y. S. All-dielectric resonant metasurfaces with a strong toroidal response. *ACS Photonics* **2018**, *5* (5), 1871-1876.

(17) Xu, S.; Sayanskiy, A.; Kupriianov, A. S.; Tuz, V. R.; Kapitanova, P.; Sun, H. B.; Han, W.; Kivshar, Y. S. Experimental observation of toroidal dipole modes in all‐dielectric metasurfaces. *Adv. Opt. Mater.* **2018**, 1801166.

(18) Jain, A.; Moitra, P.; Koschny, T.; Valentine, J.; Soukoulis, C. M. Electric and magnetic response in dielectric dark states for low loss subwavelength optical meta atoms. *Adv. Opt. Mater.* **2015**, *3* (10), 1431-1438.

(19) Campione, S.; Liu, S.; Basilio, L. I.; Warne, L. K.; Langston, W. L.; Luk, T. S.; Wendt, J. R.; Reno, J. L.; Keeler, G. A.; Brener, I.; Sinclair, M. B. Broken symmetry dielectric resonators for high quality factor fano metasurfaces. *ACS Photonics* **2016**, *3* (12), 2362-2367.
(20) Cui, C.; Zhou, C.; Yuan, S.; Qiu, X.; Zhu, L.; Wang, Y.; Li, Y.; Song, J.; Huang, Q.; Wang, Y.; Zeng, C.; Xia, J. Multiple fano resonances in symmetry-breaking silicon metasurface for manipulating light emission. *ACS Photonics* **2018**, *5* (10), 4074-4080.
(21) Tuz, V. R.; Khardikov, V. V.; Kupriianov, A. S.; Domina, K. L.; Xu, S.; Wang, H.; Sun, H. B. High-quality trapped modes in all-dielectric metamaterials. *Opt. Express* **2018**, *26* (3), 2905-2916.
(22) Wu, C.; Arju, N.; Kelp, G.; Fan, J. A.; Dominguez, J.; Gonzales, E.; Tutuc, E.; Brener, I.; Shvets, G. Spectrally selective chiral silicon metasurfaces based on infrared fano resonances. *Nat. Commun.* **2014**, *5*, 3892.
(23) Zhang, J.; MacDonald, K. F.; Zheludev, N. I. Near-infrared trapped mode magnetic resonance in an all-dielectric metamaterial. *Opt. Express* **2013**, *21* (22), 26721-8.
(24) Krasnok, A.; Tymchenko, M.; Alù, A. Nonlinear metasurfaces: A paradigm shift in nonlinear optics. *Mater. Today* **2018**, *21* (1), 8-21.
(25) Panoiu, N. C.; Sha, W. E. I.; Lei, D. Y.; Li, G. C. Nonlinear optics in plasmonic nanostructures. *J. Optics* **2018**, *20* (8), 083001.
(26) Debarre, D.; Supatto, W.; Pena, A. M.; Fabre, A.; Tordjmann, T.; Combettes, L.; Schanne-Klein, M. C.; Beaurepaire, E. Imaging lipid bodies in cells and tissues using third-harmonic generation microscopy. *Nat. Methods* **2006**, *3* (1), 47-53.
(27) Tai, S. P.; Wu, Y.; Shieh, D. B.; Chen, L. J.; Lin, K. J.; Yu, C. H.; Chu, S. W.; Chang, C. H.; Shi, X. Y.; Wen, Y. C.; Lin, K. H.; Liu, T. M.; Sun, C. K. Molecular imaging of cancer cells using plasmon-resonant-enhanced third-harmonic-generation in silver nanoparticles. *Adv. Mater.* **2007**, *19* (24), 4520-4523.
(28) Monat, C.; Grillet, C.; Collins, M.; Clark, A.; Schroeder, J.; Xiong, C.; Li, J.; O'Faolain, L.; Krauss, T. F.; Eggleton, B. J.; Moss, D. J. Integrated optical auto-correlator based on third-harmonic generation in a silicon photonic crystal waveguide. *Nat. Commun.* **2014**, *5*, 3246.
(29) Lippitz, M.; van Dijk, M. A.; Orrit, M. Third-harmonic generation from single. *Nano Lett.* **2005**, *5*, 799-802.
(30) Hentschel, M.; Utikal, T.; Giessen, H.; Lippitz, M. Quantitative modeling of the third harmonic emission spectrum of plasmonic nanoantennas. *Nano Lett.* **2012**, *12* (7), 3778-82.
(31) Reinhold, J.; Shcherbakov, M. R.; Chipouline, A.; Panov, V. I.; Helgert, C.; Paul, T.; Rockstuhl, C.; Lederer, F.; Kley, E. B.; Tünnermann, A.; Fedyanin, A. A.; Pertsch, T. Contribution of the magnetic resonance to the third harmonic generation from a fishnet metamaterial. *Phy. Rev. B* **2012**, *86* (11), 115401.
(32) Lassiter, J. B.; Chen, X.; Liu, X.; Ciracì, C.; Hoang, T. B.; Larouche, S.; Oh, S.-H.; Mikkelsen, M. H.; Smith, D. R. Third-harmonic generation enhancement by film-coupled plasmonic stripe resonators. *ACS Photonics* **2014**, *1* (11), 1212-1217.
(33) Metzger, B.; Schumacher, T.; Hentschel, M.; Lippitz, M.; Giessen, H. Third harmonic mechanism in complex plasmonic fano structures. *ACS Photonics* **2014**, *1* (6), 471-476.
(34) Zhai, W. C.; Qiao, T. Z.; Cai, D. J.; Wang, W. J.; Chen, J. D.; Chen, Z. H.; Liu, S. D. Anticrossing double fano resonances generated in metallic/dielectric hybrid nanostructures using nonradiative anapole modes for enhanced nonlinear optical effects. *Opt. Express* **2016**, *24* (24), 27858-27869.
(35) Aouani, H.; Rahmani, M.; Navarro-Cia, M.; Maier, S. A. Third-harmonic-upconversion enhancement from a single semiconductor nanoparticle coupled to a plasmonic antenna. *Nat. Nanotechnol.* **2014**, *9* (4), 290-4.

(36) Shibanuma, T.; Grinblat, G.; Albella, P.; Maier, S. A. Efficient third harmonic generation from metal-dielectric hybrid nanoantennas. *Nano Lett.* **2017**, *17* (4), 2647-2651.

(37) Staude, I.; Schilling, J. Metamaterial-inspired silicon nanophotonics. *Nat. Photonics* **2017**, *11* (5), 274-284.

(38) Shcherbakov, M. R.; Neshev, D. N.; Hopkins, B.; Shorokhov, A. S.; Staude, I.; Melik-Gaykazyan, E. V.; Decker, M.; Ezhov, A. A.; Miroshnichenko, A. E.; Brener, I.; Fedyanin, A. A.; Kivshar, Y. S. Enhanced third-harmonic generation in silicon nanoparticles driven by magnetic response. *Nano Lett.* **2014**, *14* (11), 6488-92.

(39) Shorokhov, A. S.; Melik-Gaykazyan, E. V.; Smirnova, D. A.; Hopkins, B.; Chong, K. E.; Choi, D. Y.; Shcherbakov, M. R.; Miroshnichenko, A. E.; Neshev, D. N.; Fedyanin, A. A.; Kivshar, Y. S. Multifold enhancement of third-harmonic generation in dielectric nanoparticles driven by magnetic fano resonances. *Nano Lett.* **2016**, *16* (8), 4857-61.

(40) Smirnova, D. A.; Khanikaev, A. B.; Smirnov, L. A.; Kivshar, Y. S. Multipolar third-harmonic generation driven by optically induced magnetic resonances. *ACS Photonics* **2016**, *3* (8), 1468-1476.

(41) Wang, L.; Kruk, S.; Xu, L.; Rahmani, M.; Smirnova, D.; Solntsev, A.; Kravchenko, I.; Neshev, D.; Kivshar, Y. Shaping the third-harmonic radiation from silicon nanodimers. *Nanoscale* **2017**, *9* (6), 2201-2206.

(42) Corcoran, B.; Monat, C.; Grillet, C.; Moss, D. J.; Eggleton, B. J.; White, T. P.; O'Faolain, L.; Krauss, T. F. Green light emission in silicon through slow-light enhanced third-harmonic generation in photonic-crystal waveguides. *Nat. Photonics* **2009**, *3* (4), 206-210.

(43) Shcherbakov, M. R.; Shorokhov, A. S.; Neshev, D. N.; Hopkins, B.; Staude, I.; Melik-Gaykazyan, E. V.; Ezhov, A. A.; Miroshnichenko, A. E.; Brener, I.; Fedyanin, A. A.; Kivshar, Y. S. Nonlinear interference and tailorable third-harmonic generation from dielectric oligomers. *ACS Photonics* **2015**, *2* (5), 578-582.

(44) Yang, Y.; Wang, W.; Boulesbaa, A.; Kravchenko, II; Briggs, D. P.; Puretzky, A.; Geohegan, D.; Valentine, J. Nonlinear fano-resonant dielectric metasurfaces. *Nano Lett.* **2015**, *15* (11), 7388-93.

(45) Tong, W.; Gong, C.; Liu, X.; Yuan, S.; Huang, Q.; Xia, J.; Wang, Y. Enhanced third harmonic generation in a silicon metasurface using trapped mode. *Opt. Express* **2016**, *24* (17), 19661-70.

(46) Chen, S.; Rahmani, M.; Li, K. F.; Miroshnichenko, A.; Zentgraf, T.; Li, G.; Neshev, D.; Zhang, S. Third harmonic generation enhanced by multipolar interference in complementary silicon metasurfaces. *ACS Photonics* **2018**, *5* (5), 1671-1675.

(47) Shcherbakov, M. R.; Werner, K.; Fan, Z.; Talisa, N.; Chowdhury, E.; Shvets, G. Photon acceleration and tunable broadband harmonics generation in nonlinear time-dependent metasurfaces. *Nat. Commun.* **2019**, *10* (1), 1345.

(48) Okhlopkov, K. I.; Shafirin, P. A.; Ezhov, A. A.; Orlikovsky, N. A.; Shcherbakov, M. R.; Fedyanin, A. A. Optical coupling between resonant dielectric nanoparticles and dielectric nanowires probed by third harmonic generation microscopy. *ACS Photonics* **2018**, *6* (1), 189-195.

(49) Chen, S.; Li, Z.; Liu, W.; Cheng, H.; Tian, J. From single-dimensional to multidimensional manipulation of optical waves with metasurfaces. *Adv. Mater.* **2019**, *31* (16), 1802458.

(50) Koshelev, K.; Tang, Y.; Li, K.; Choi, D.-Y.; Li, G.; Kivshar, Y. Nonlinear metasurfaces governed by bound states in the continuum. *ACS Photonics* **2019**, doi.org/10.1021/acsphotonics.9b00700.

(51) Yang, Y.; Kravchenko, II; Briggs, D. P.; Valentine, J. All-dielectric metasurface analogue of electromagnetically induced transparency. *Nat. Commun.* **2014**, *5*, 5753.

(52) Fedotov, V. A.; Papasimakis, N.; Plum, E.; Bitzer, A.; Walther, M.; Kuo, P.; Tsai, D. P.; Zheludev, N. I. Spectral collapse in ensembles of metamolecules. *Phys. Rev. Lett.* **2010**, *104* (22), 223901.

(53) Galli, M.; Portalupi, S. L.; Belotti, M.; Andreani, L. C.; O'Faolain, L.; Krauss, T. F. Light scattering and fano resonances in high-q photonic crystal nanocavities. *Appl. Phys. Lett.* **2009**, *96*, 071101.

(54) Yanik, A. A.; Cetin, A. E.; Huang, M.; Artar, A.; Mousavi, S. H.; Khanikaev, A.; Connor, J. H.; Shvets, G.; Altug, H. Seeing protein monolayers with naked eye through plasmonic fano resonances. *Proc. Natl. Acad. Sci. USA* **2011**, *108* (29), 11784-9.

(55) Lim, W. X.; Manjappa, M.; Pitchappa, P.; Singh, R. Shaping high-q planar fano resonant metamaterials toward futuristic technologies. *Adv. Opt. Mater.* **2018**, 1800502.

(56) Zhou, C.; Liu, G.; Ban, G.; Li, S.; Huang, Q.; Xia, J.; Wang, Y.; Zhan, M. Tunable fano resonator using multilayer graphene in the near-infrared region. *Appl. Phys. Lett.* **2018**, *112* (10), 101904.

(57) Wu, P. C.; Liao, C. Y.; Savinov, V.; Chung, T. L.; Chen, W. T.; Huang, Y. W.; Wu, P. R.; Chen, Y. H.; Liu, A. Q.; Zheludev, N. I.; Tsai, D. P. Optical anapole metamaterial. *ACS Nano* **2018**, *12* (2), 1920-1927.

(58) Koshelev, K.; Lepeshov, S.; Liu, M.; Bogdanov, A.; Kivshar, Y. Asymmetric metasurfaces with high-q resonances governed by bound states in the continuum. *Phys. Rev. Lett.* **2018**, *121* (19), 193903.

(59) Xu, L.; Zangeneh Kamali, K.; Huang, L.; Rahmani, M.; Smirnov, A.; Camacho‐Morales, R.; Ma, Y.; Zhang, G.; Woolley, M.; Neshev, D.; Miroshnichenko, A. E. Dynamic nonlinear image tuning through magnetic dipole quasi‐bic ultrathin resonators. *Adv. Sci.* **2019**, 1802119.

(60) Baranov, D. G.; Zuev, D. A.; Lepeshov, S. I.; Kotov, O. V.; Krasnok, A. E.; Evlyukhin, A. B.; Chichkov, B. N. All-dielectric nanophotonics: The quest for better materials and fabrication techniques. *Optica* **2017**, *4* (7), 814.

(61) Vabishchevich, P. P.; Liu, S.; Sinclair, M. B.; Keeler, G. A.; Peake, G. M.; Brener, I. Enhanced second-harmonic generation using broken symmetry iii–v semiconductor fano metasurfaces. *ACS Photonics* **2018**, *5* (5), 1685-1690.

(62) Liu, H.; Guo, C.; Vampa, G.; Zhang, J. L.; Sarmiento, T.; Xiao, M.; Bucksbaum, P. H.; Vučković, J.; Fan, S.; Reis, D. A. Enhanced high-harmonic generation from an all-dielectric metasurface. *Nat. Phys.* **2018**, *14*, 1006–1010.

(63) Makarov, S. V.; Petrov, M. I.; Zywietz, U.; Milichko, V.; Zuev, D.; Lopanitsyna, N.; Kuksin, A.; Mukhin, I.; Zograf, G.; Ubyivovk, E.; Smirnova, D. A.; Starikov, S.; Chichkov, B. N.; Kivshar, Y. S. Efficient second-harmonic generation in nanocrystalline silicon nanoparticles. *Nano Lett.* **2017**, *17* (5), 3047-3053.

(64) Vampa, G.; Ghamsari, B. G.; Siadat Mousavi, S.; Hammond, T. J.; Olivieri, A.; Lisicka-Skrek, E.; Naumov, A. Y.; Villeneuve, D. M.; Staudte, A.; Berini, P.; Corkum, P. B. Plasmon-enhanced high-harmonic generation from silicon. *Nat. Phys.* **2017**, *13* (7), 659-662.

(65) Palik, E. D. Handbook of optical constants of solids. *Academic Press, San Diego, CA*, 1998.

(66) Ban, G.; Gong, C.; Zhou, C.; Li, S.; Barille, R.; Liu, X.; Wang, Y. Fano-resonant silicon photonic crystal slab for efficient third-harmonic generation. *Opt. Lett.* **2019**, *44* (1), 126-129.

# Supporting Information

# Resonant Asymmetric Dielectric Metasurface for Boosting Third-Harmonic Generation

Chaobiao Zhou,[†,§,⊥] Shiyu Li,[†] Cheng Gong,[§] Yi Wang,[*,†] Xiaojun Liu,[*,§] and Mingsheng Zhan[*,†,§]

[†]*Wuhan National Laboratory for Optoelectronics, Huazhong University of Science and Technology, Wuhan, Hubei, 430074, China*

[§]*State Key Laboratory of Magnetic Resonance and Atomic and Molecular Physics, Wuhan Institute of Physics and Mathematics, Chinese Academy of Sciences, Wuhan, Hubei, 430071, China*

[⊥]*College of Mechanical and Electronic Engineering, Guizhou Minzu University, Guiyang 550025, China*

*E-mail: ywangwnlo@mail.hust.edu.cn.
*E-mail: xjliu@wipm.ac.cn.
*E-mail: mszhan@wipm.ac.cn.

1. **Derivation of the dependence of the Q-factor on the asymmetric parameter α**

For a resonant state with a complex frequency $\omega = \omega_0 - i\gamma/2$, The mode inverse radiation lifetime $\gamma$ can be expressed as[1]

$$\gamma = \frac{Ac \sum_{i=x,y} \left( |E_{i,+1}|^2 + |E_{i,-1}|^2 \right)\Big|_{z \in S}}{\int_V dV \, \varepsilon |\vec{E}|^2}, \qquad (1)$$

where $A$ is the surface area of a unit cell, $E_{i,+1}$ and $E_{i,-1}$ are the $x$- or $y$- component of the electric field in the far-field of the resonant state along $+z$ and $-z$ directions, respectively, $S$ is boundaries lying in the far-field enclosed the structures, the volume integration calculates the electric field of the resonant state between the boundaries. Since the resonance we study has less energy leakage, the first order perturbation theory $\int_V dV \, \varepsilon |\vec{E}|^2 = 1$ can be applied to simplify Eq. (1) to

$$\gamma = Ac \sum_{i=x,y} \left( |E_{i,+1}|^2 + |E_{i,-1}|^2 \right)\Big|_{z \in S}. \qquad (2)$$

Further applying the Lippmann-Schwinger equation and perturbation theory to the resonant state, $\gamma$ is written as

$$\gamma = c \sum_{i=x,y} |D_i|^2, \tag{3}$$

where coupling amplitudes $D_i$ are defined as $D_i = -\frac{\omega_0}{\sqrt{2Ac}} \int d\vec{r}' \left[\varepsilon(\omega_0, \vec{r}') - 1\right] E_i(\vec{r}') e^{ik_0 z'}$, $i = x, y$.

Expanding the function $e^{ik_0 z'}$ into the Taylor series, we obtain

$$D_x = -\frac{k_0}{\sqrt{2A}} \left(p_x - \frac{m_y}{c} + \frac{ik_0}{6} Q_{zx}\right), \tag{4a}$$

$$D_y = -\frac{k_0}{\sqrt{2A}} \left(p_y + \frac{m_x}{c} + \frac{ik_0}{6} Q_{yz}\right), \tag{4b}$$

where $\vec{p}, \vec{m}, \vec{Q}$ are the electric dipole, magnetic dipole and electric quadrupole moments in the irreducible representations, $k_0$ is the incident wave vector in the free space. Other higher-order multipoles are negligible due to the satisfaction of $hk_0 < 1$, where $h$ is the height of the structures.[2] Because of the broken mirror symmetry along the y-axis, $E_x$ and $E_y$ are classified as odd and even, respectively, as verified in Figure S1. Therefore, only $D_y$ accounts for $\gamma$, given that the integration of $E_x$ equals 0. Moreover, $m_x$ and $Q_{yz}$ equal 0 owing to the up-down symmetry $E_y(-z) = E_y(z)$. Finally, $\gamma$ can be written as

$$\gamma = \frac{k_0^2 c}{2A} |p_y|^2. \tag{5}$$

As denoted with gray arrows in the $E_y$ profile of Figure. S1a, $p_y$ can be seen as the uncompensated dipole moments with opposite directions supported by the two halves of the structure with respect to the y-axis, satisfying $p_y = \Delta S / S \, p_0 = \alpha p_0$, where $\Delta S$ is the area of the cut-off part, $S$ is the area of the symmetric nanopillar, $\alpha = \Delta S / S$ is the asymmetry parameter, $p_0$ is the electric dipole moment in the right half of the nanopillar. Finally, the radiative quality factor $Q_{rad} = \omega_0 / \gamma$ of the resonant state can be found as

$$Q_{rad} = \frac{2A}{k_0} |p_0|^{-2} \alpha^{-2}, \tag{6}$$

demonstrating an inverse quadratic trend as shown in Figure. 4.

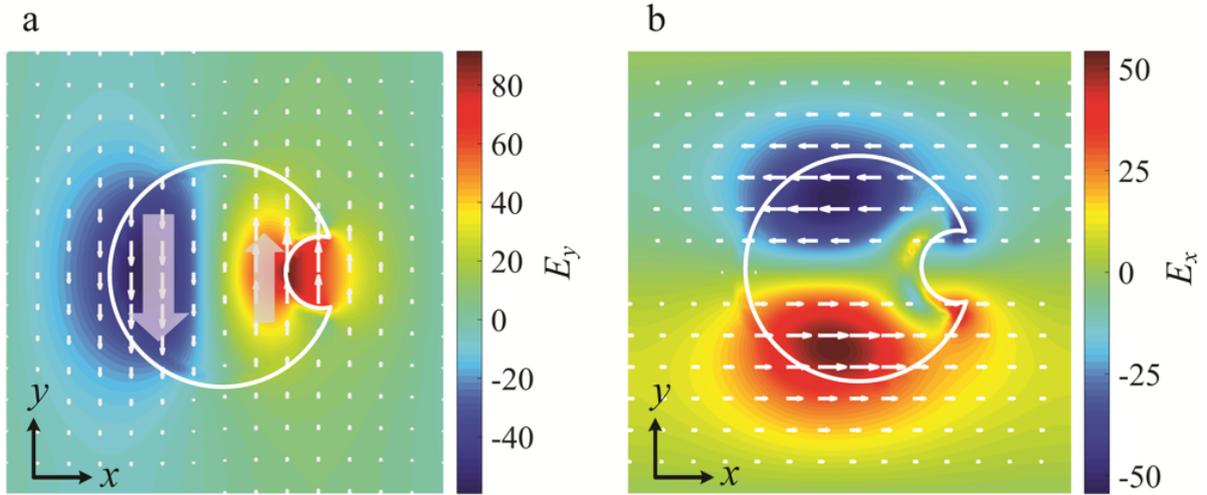

**Figure S1:** Calculated electric fields in the *x-y* plane for $E_y$ and $E_x$ components, respectively.

## 2. The magnetic field distribution in *x-z* plane

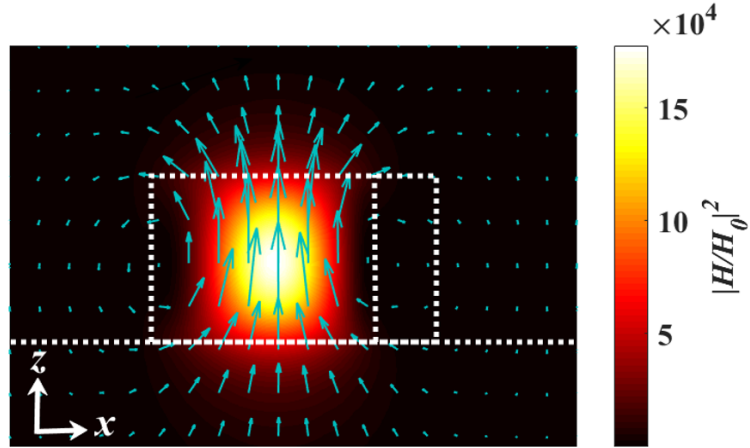

**Figure S2.** The magnetic field distribution at resonance wavelength $\lambda_{Res} = 1341$ nm at the *x-z* plane. It is clear that the magnetic dipole moment oscillates along the *z*- axis.

## 3. The power dependence of THG in metasurface

In order to gain further insight into the nature of nonlinear processes in the metasurface, we investigate the power dependence of the THG response. As shown in Figure S3a, in the low power range, the THG intensity increases as the optical pump power increases. The saturation phenomenon is observed at the high power range, which stems from the two-photon absorption (free carrier generation) in the bulk silicon substrate.[3-4] In the inset of Figure S3a, we show the THG spectra with varying low pump intensity, and the THG is observed under exceptionally low pump power. In the Figure S3b, we fit the experimental data in perturbative regime with a third-order power function, and the output energy of THG shows a near-cubic (slope = 2.75) dependence on the pumping power, which is close to the theoretical value (slope = 3) for third-order optical nonlinearity. The deviation may be induced by the unstable power of femtosecond laser and some other nonlinear effects, such as the nonlinear Kerr and TPA effects, free carrier absorption, and the thermo-optical effect.

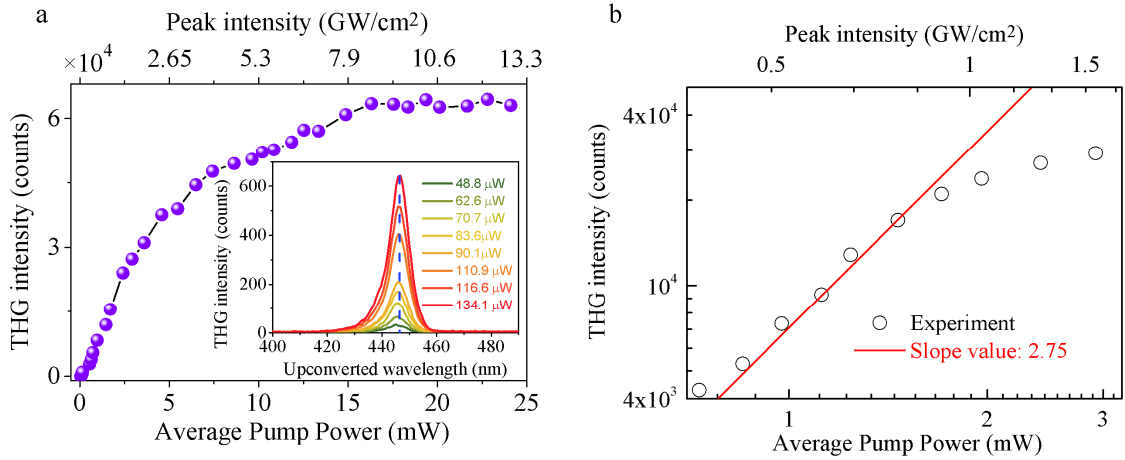

**Figure S3.** a) The measured THG intensity for silicon metasurface as a function of pump power intensity when the pump pulse is centered at 1341 nm, with inset showing the THG spectra in low pump power region. b) Log plot of the third harmonic power as a function of pump power and peak pump intensity. The black circles indicate the measured data, and the red line is a numerical fit to the data with a third-order power function.

## 4. The diffraction of THG in the resonant metasurface

We calculate the third harmonic diffraction from the Fano-resonant metasurface using a finite-difference time-domain (FDTD) solver. The simulation method is used by referencing literatures.[4-6] The dielectric constants of Si and $SiO_2$ are adopted from the Palik.[7] The parameters of light source keep pace with the employed femtosecond laser in experiment. Figure S4a shows the schematic with definition of the far-field coordinates ($\theta$, $\varphi$). The incident electric field with the center wavelength 1341 nm is propagating along the *z*- axis and is polarized along the *y*-axis, the asymmetric nanodisk is in the *x-y* plane, and the structure parameters match with the main text. The simulated results are shown in Figure S4b, which indicates that there exist first and second order third-harmonic diffractions in our design.

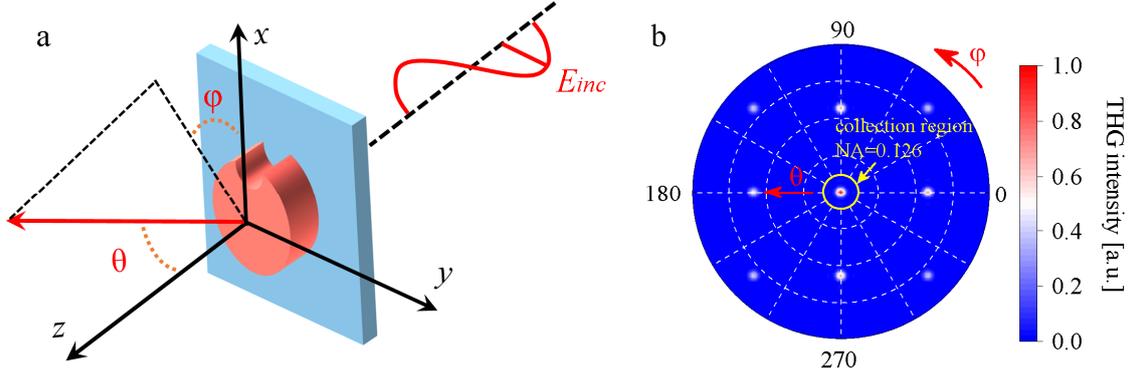

**Figure S4.** a) Schematic of definition of far-field coordinates ($\theta$, $\varphi$). b) Simulated THG far-field radiation for different coordinates ($\theta$, $\varphi$), the marked collection region corresponds to the case of N. A. = 0.126 used in our experiment in the inset.

In addition, the diffraction angle $\theta$ depends on wavelength $\lambda$ of the light and period $P$ of the metasurface through the following formula:[8]

$$\sqrt{n_x^2 + n_y^2}\,\lambda = P\sin\theta, \tag{7}$$

$n_x$, $n_y$ are diffraction orders in the $x$-, $y$- directions, respectively.

The theoretical results calculated by Eq. (7) and the simulated values (FDTD) are listed in the following Tab. S1. They match well with each other. In our experiment, a numerical aperture (N. A. = 0.126) objective lens is used to collect THG signal. Thus, only the zero order diffraction power is collected, the collection region is shown in Figure S4b, and the collected power is about 12.9%.

**Table S1.** The diffraction results from theoretical calculation and FDTD simulation.

| Diffraction order | Diffraction angle $\theta$ for theory | Diffraction angle $\theta$ for simulation | Power (%) |
|---|---|---|---|
| 0th | 0º | 0º | 12.9 |
| 1st | 33.79º | 33.73º | 46.9 |
| 2nd | 51.87º | 51.75º | 40.2 |

## Supplementary References:

(1) Koshelev, K.; Lepeshov, S.; Liu, M.; Bogdanov, A.; Kivshar, Y. Asymmetric metasurfaces with high-q resonances governed by bound states in the continuum. *Phys. Rev. Lett.* **2018**, *121* (19), 193903.

(2) Evlyukhin, A. B.; Fischer, T.; Reinhardt, C.; Chichkov, B. N. Optical theorem and multipole scattering of light by arbitrarily shaped nanoparticles. *Phys. Rev. B* **2016**, *94* (20), 205434.

(3) Shcherbakov, M. R.; Neshev, D. N.; Hopkins, B.; Shorokhov, A. S.; Staude, I.; Melik-Gaykazyan, E. V.; Decker, M.; Ezhov, A. A.; Miroshnichenko, A. E.; Brener, I.; Fedyanin, A. A.; Kivshar, Y. S. Enhanced third-harmonic generation in silicon nanoparticles driven by magnetic response. *Nano Lett.* **2014**, *14* (11), 6488-92.

(4) Yang, Y.; Wang, W.; Boulesbaa, A.; Kravchenko, II; Briggs, D. P.; Puretzky, A.; Geohegan, D.; Valentine, J. Nonlinear fano-resonant dielectric metasurfaces. *Nano Lett.* **2015**, *15* (11), 7388-93.

(5) Zhai, W. C.; Qiao, T. Z.; Cai, D. J.; Wang, W. J.; Chen, J. D.; Chen, Z. H.; Liu, S. D. Anticrossing double fano resonances generated in metallic/dielectric hybrid nanostructures using nonradiative anapole modes for enhanced nonlinear optical effects. *Opt. Express* **2016**, *24* (24), 27858-27869.

(6) Löchner, F. J. F.; Fedotova, A. N.; Liu, S.; Keeler, G. A.; Peake, G. M.; Saravi, S.; Shcherbakov, M. R.; Burger, S.; Fedyanin, A. A.; Brener, I.; Pertsch, T.; Setzpfandt, F.; Staude, I. Polarization-dependent second harmonic diffraction from resonant gaas metasurfaces. *ACS Photonics* **2018**, *5* (5), 1786-1793.

(7) Palik, E. D. Handbook of optical constants of solids. *Academic Press, San Diego, CA*, 1998.

(8) Nshii, C. C.; Vangeleyn, M.; Cotter, J. P.; Griffin, P. F.; Hinds, E. A.; Ironside, C. N.; See, P.; Sinclair, A. G.; Riis, E.; Arnold, A. S. A surface-patterned chip as a strong source of ultracold atoms for quantum technologies. *Nat. Nanotechnol.* **2013**, *8* (5), 321-324.